# Effects of surface structure deformation on static friction at fractal interfaces

D. A. H. HANAOR*, Y. GAN* and I. EINAV*

The evolution of fractal surface structures with flattening of asperities was investigated using isotropically roughened aluminium surfaces loaded in compression. It was found that asperity amplitude, mean roughness and fractal dimension decrease through increased compressive stress and number of loading events. Of the samples tested, surfaces subjected to an increased number of loading events exhibited the most significant surface deformation and were observed to exhibit higher levels of static friction at an interface with a single-crystal flat quartz substrate. This suggests that the frequency of grain reorganisation events in geomaterials plays an important role in the development of intergranular friction. Fractal surfaces were numerically modelled using Weierstrass–Mandelbrot-based functions. From the study of frictional interactions with rigid flat opposing surfaces it was apparent that the effect of surface fractal dimension is more significant with increasing dominance of adhesive mechanisms.

## INTRODUCTION

Across a range of scales, from clay platelets to fault gouges, static friction at granular interfaces governs geomechanical processes such as the overall strength and deformation of granular materials. For most purposes, frictional phenomena in geomaterials are described at a continuum level, applying empirically determined friction coefficients at the macroscopic scale. Such approaches lack the ability to account for the evolution of surface structures through wear and deformation of surface asperities at the grain scale, with significant consequences on bulk behaviour (Cavarretta *et al.*, 2010). Understanding the micromechanical origins of frictional phenomena and the development of surface structures is of great importance to enable meaningful modelling of time-dependent intergranular forces in geomaterials.

Interfacial frictional phenomena arise from the interaction of contacting micro-asperities. Although the significance of roughness on parameters of intergranular friction has been reported for specific systems (Sadrekarimi & Olson, 2011), to date no comprehensive relationship has been established between roughness descriptors and frictional phenomena. Conventional studies into contact mechanics generally assume single distributions of asperity heights, often with spherical contacts (Greenwood & Williamson, 1966). These models and associated descriptors of roughness are inadequate for the interpretation of interfacial resistance to shear and indeed, under divergent conditions, increasing contact roughness may result in either higher or lower values of static friction (Persson, 2006, 2007). Much of the complexity of surfaces arises from their fractal structures, with self-similar features present at ever-finer scales (Sammis & Biegel, 1989; Majumdar & Tien, 1990; Yan & Komvopoulos, 1998). Considering the fractal dimension of surfaces is of great importance in the simulation of granular materials (as shown in Fig. 1), where the ability to predict frictional forces arising from multiscale asperity structures would enable meaningful continuum-scale modelling. In recent years, contact mechanics of fractal surfaces has attracted increasing attention and the fractal nature of surfaces is emerging as a key factor governing the mechanical, physical and chemical properties of materials (Buzio *et al.*, 2003; Goedecke *et al.*, 2012).

At a molecular scale, asperity interactions are inadequate to explain frictional forces and mechanisms of adhesion are more apt to describe resistance to shear (Li & Kim, 2008). Atomic frictional phenomena differ from

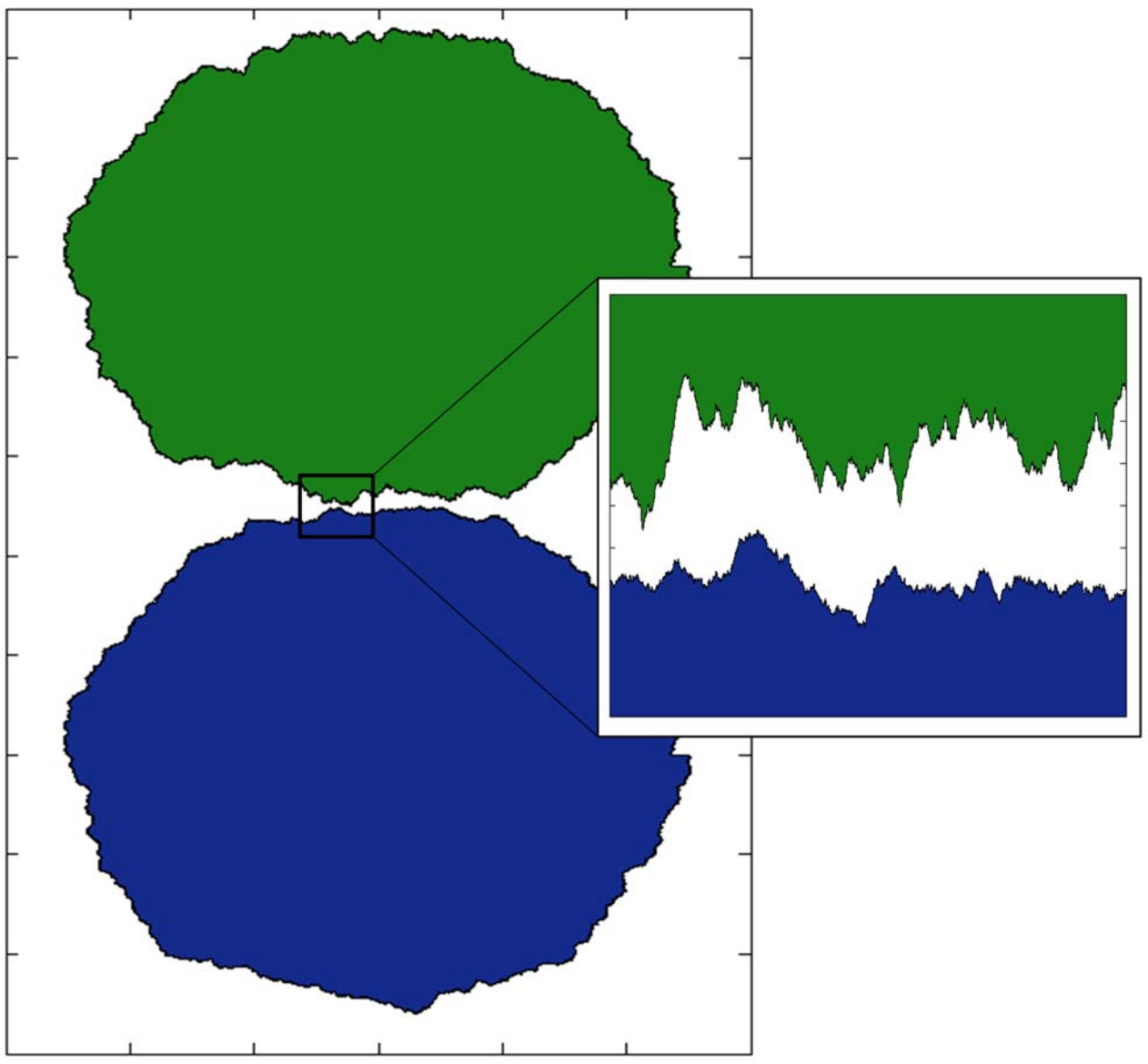

**Fig. 1.** Two-dimensional simulation of quasi-spherical grains with magnified fractal surface structures shown in inset

*Particles and Grains Laboratory, School of Civil Engineering, University of Sydney, NSW, Australia

classical Amonton–Coulomb friction. As with the Terzaghi–Bowden–Tabor adhesive model of friction, asperity contact points at a molecular level can be considered as cold-welded junctions (Bowden & Tabor, 2001; Patra *et al.*, 2008). The number of contacting asperities, their geometric distribution and their individual loading conditions are significantly affected by the fractal dimension of surface structures. Owing to surface adhesion, asperity contacts may exhibit nano-scale localised negative frictional behaviour whereby the local frictional force decreases as two points on opposing surfaces approach each other (Deng *et al.*, 2012).

The multiscale micromechanics of friction have been explored in several publications (Boitnott *et al.*, 1992; Harrison *et al.*, 1992). At the finest scales, frictional stress $\tau_f$ is described by a load-dependent component, governed by the coefficient of molecular scale friction $\alpha$ and a material-interface-dependent adhesive shear stress $\tau_0$ (Li & Kim, 2008)

$$\tau_f = \tau_0 + \alpha P \quad (1)$$

where $P$ is the contact pressure, which is the sum of applied and capillary-induced components. The relative significance of the material- and load-dependent components is strongly influenced by the contact profile and fractality of surface structures.

The work reported here experimentally investigated the evolution of fractal geometries with asperity flattening and employed concepts of atomic-scale friction to qualitatively evaluate the static frictional behaviour of computer-generated fractal surfaces on a point-by-point basis to understand the significance of fractal dimensions on the frictional behaviour of multiscale structures.

## METHOD

### *Experimental measurements of surface structure evolution*

To evaluate an evolving asperity structure, discs of 25 mm diameter and 3 mm thickness were prepared from aluminium Al-2011 and homogenous isotropic surface roughness was imparted through the high-velocity spraying of 250–300 μm size glass beads. Asperity deformation was achieved by compressing pairs of roughened surfaces against each other using a hydraulic uniaxial press at constant loads for a duration of 10 min with applied stresses of 20·37, 40·74, 61·11, 101·86, 152·78 and 203·72 MPa. Although a nano-scale brittle oxide layer is likely to be present on the material surface and exhibit breakage, asperity deformation is assumed to be predominantly plastic. Applied stresses were found to be sufficient to impart surface-structure flattening without the onset of bulk deformation. In a separate series of samples, compression at 152·78 MPa was applied in multiple events, with samples rotated with respect to one another between events. This was designed to give rise to new asperity contact distributions.

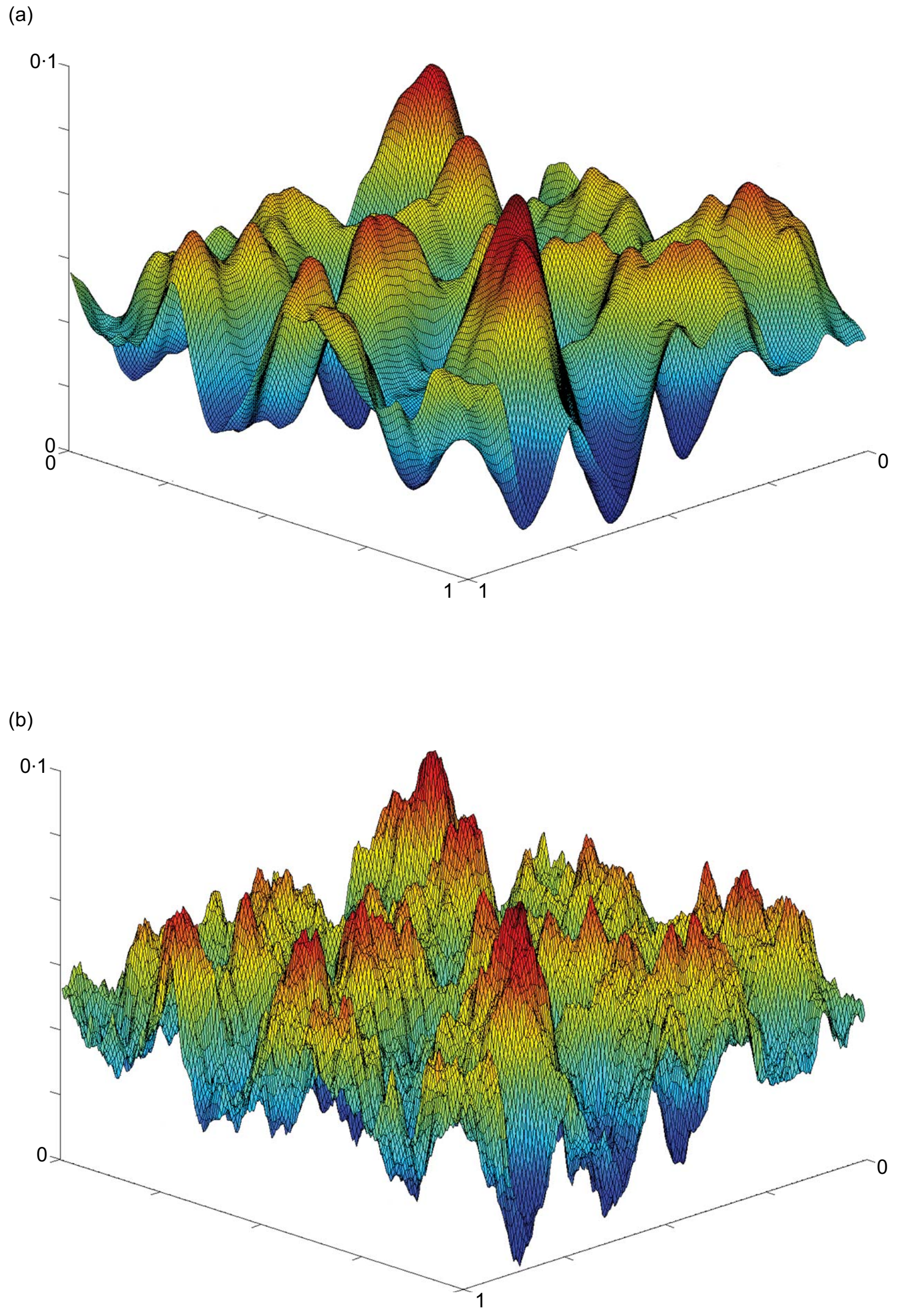

**Fig. 2.** (a) Initial non-fractal surface from which the fractal surface shown in (b) with fractal dimension $D$ = 2·5 was constructed

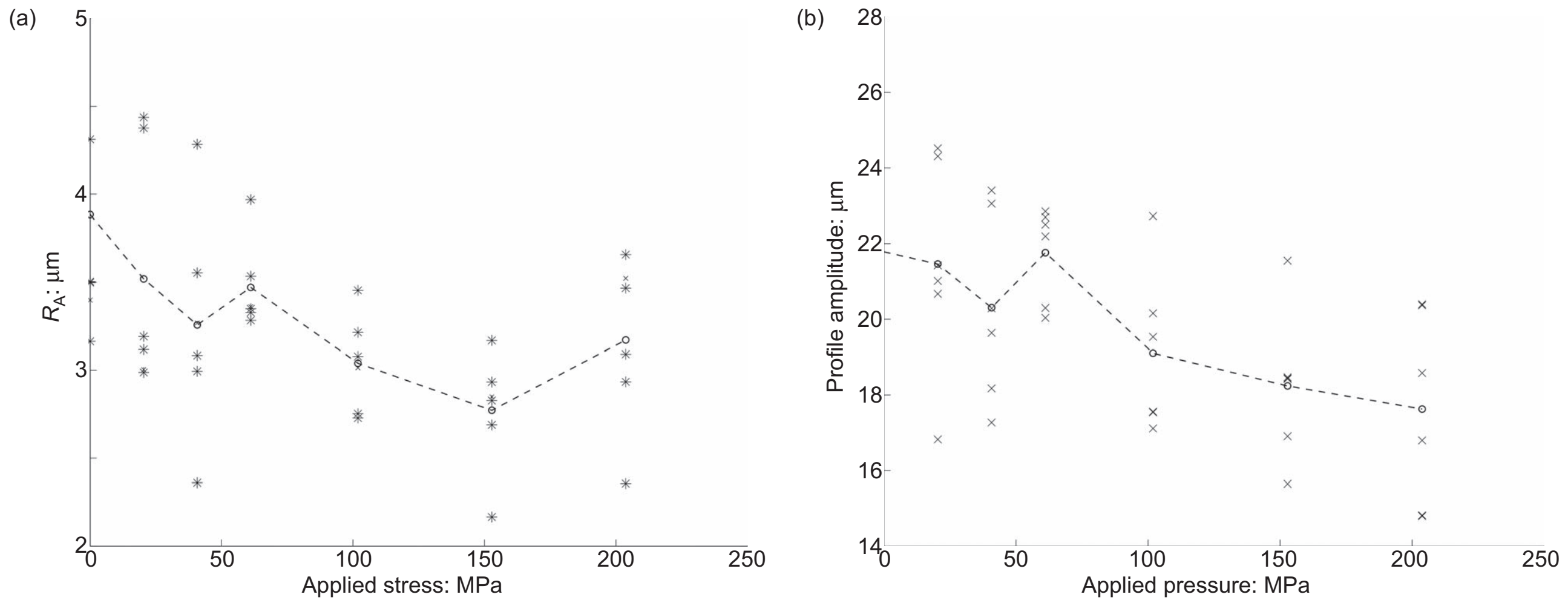

**Fig. 3.** Variation of (a) mean roughness and (b) profile amplitude with compressive stress applied over the apparent specimen area of contact

Surface structures were evaluated using a stylus profilometer (Tencor P-11) with a 2 μm radius stylus point. Scans were carried out over 2000 μm lengths at 50 μm/s with data acquisition at 200 Hz, giving a maximal lateral resolution of 0·25 μm. Scans were repeated five times per specimen. These parameters were found to be suitable for obtaining representative surface profiles. Fractal dimensions of one-dimensional profiles were evaluated using methods similar to those used elsewhere, utilising the log–log change of apparent profile length relative to the resolution of measured points (Hasegawa *et al.*, 1996; Sun & Xu, 2005). As fractal dimensions were recorded from linear scans, they lie in the range 1–2. The resolution of measured points was decreased by selecting data points with increasing separation, from a point spacing of 0·25 μm (full resolution) to 250 μm.

Subsequent to surface deformation, static frictional interactions between roughened aluminium surfaces and a rigid flat surface were measured using a single crystal α-quartz [0001] substrate, which is considered as an atomically flat rigid counter surface. Frictional measurements were based on gravity-driven slipping of aluminium discs on the substrate. This was carried out without the application of an additional normal load beyond the normal component of the weight of the specimen (4·3 ± 0·3 g). In this experimental setup, normal loads are sufficiently low to avoid abrasion of surfaces and adhesion-based friction dominates as asperity–asperity interactions are essentially absent even at nanometre scales. To minimise humidity effects, physically adsorbed water was removed by drying the samples and quartz substrates at 110°C prior to frictional experimentation. Friction angles were increased using a screw-driven tilting hinged stage to determine the maximal achievable angle of adhesion of the aluminium discs to the quartz substrates. Using this approach was found to give more consistent results than using minimum angles of slip, as contact ageing effects are thus minimised. Friction measurements were repeated five times and exhibited a variance range of around $\pm 20\%$.

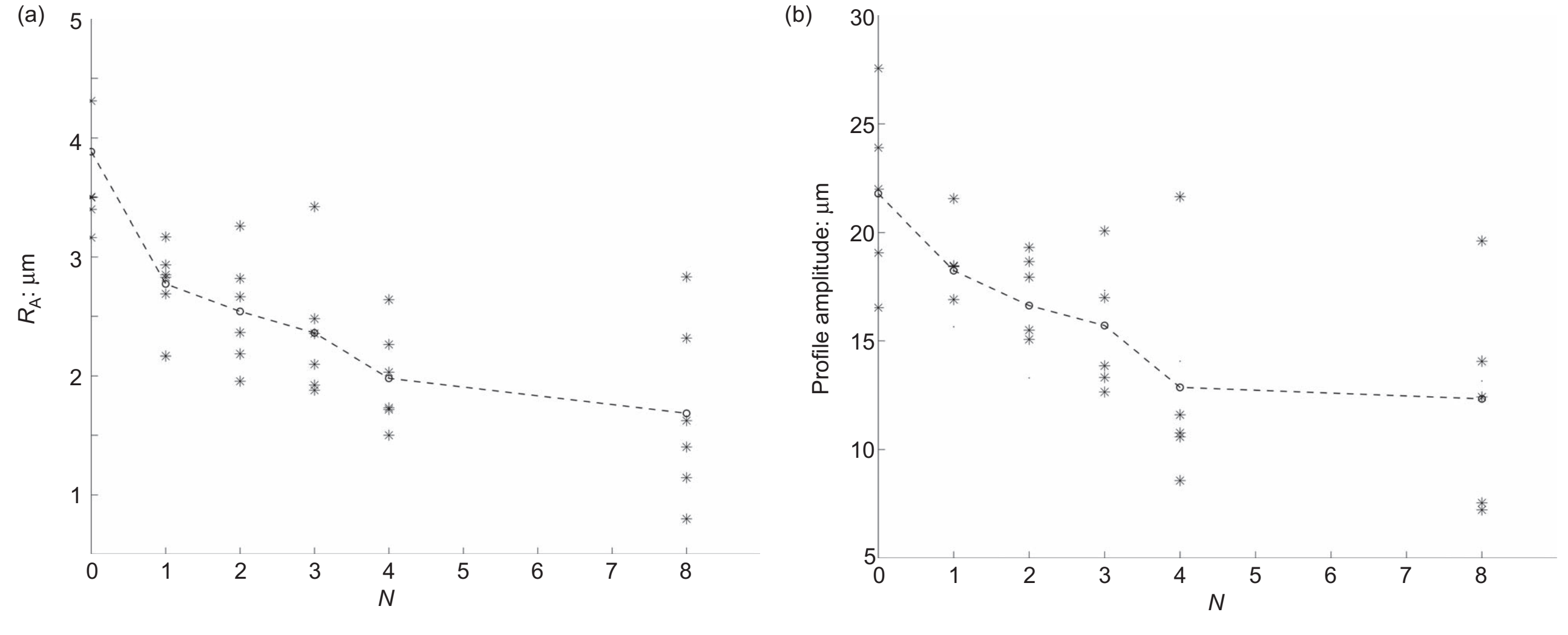

**Fig. 4.** Variation of (a) mean roughness and (b) profile amplitude with number of compression events *N*

### *Simulation of fractal surface structures*

Series of quasi-random three-dimensional fractal surfaces were isotropically constructed where the surface profile height $z$ is a function of planar coordinates $x$ and $y$ following a variant of the Weierstrass–Mandelbrot function shown in equation (2) (Berry & Lewis, 1980; Yan & Komvopoulos, 1998).

$$z(x,y) = \left(\frac{\ln\gamma}{M}\right)^{1/2} L^{(3-D)} G^{(D-2)}$$

$$\sum_{m=1}^{M}\sum_{0}^{n_{\max}} \gamma^{(D-3)n}\left\{\left[\cos\phi_{m,n} - \cos\frac{2\pi\gamma^{n}(x^2+y^2)^{1/2}}{L}\right.\right. \tag{2}$$

$$\left.\left.\cos\left(\tan^{-1}\left(\frac{y}{x}\right) - \frac{\pi m}{M}\right) + \phi_{m,n}\right]\right\}$$

As shown in Fig. 2, following this approach, fractal surfaces were constructed on the basis of a surface with randomly positioned macro-asperities where $L$ is the length factor indicative of the highest scale asperity spacing, $G$ is the fractal roughness, $D$ is the fractal dimension (which is between 2 and 3 in three dimensions), $\gamma$ is the density of frequencies used in generating the surface, $M$ is the number of superimposed ridges used to construct the surface and $\varphi_{\mathrm{m},n}$ is a randomised set of phase angles used to generate stochasticity in the surface. In the simulations reported here, non-dimensionalised values of $\gamma = 1{\cdot}5$, $M = 10$, $n_{\max} = 100$, $G = 5$ and $L = 50$ were set for a simulated surface of $150 \times 150$ points, which was subsequently scaled to give a constant amplitude.

Static friction of simulated surfaces in contact with a rigid flat were evaluated by considering each point on a fractal surface as an elastically deforming square box-element with Poisson's ratio $v = 0$ without shearing between elements. This simplified approach was taken to isolate the effects of the surface profile independently of the material properties. A constant non-dimensionalised load was applied in simulations, which imparted surface deformations in the range 0·2–0·5 of the surface profile amplitude, a similar deformation to that exhibited by experimental samples. The friction of each element with the opposing surface was governed by equation (1) with an applied limit $\tau_{\max}$ to the frictional stress value of an individual contact point representing asperity yielding.

## RESULTS

### *Experimental results*

With increasing compressive stress, surface amplitude and arithmetic mean roughness ($R_{\mathrm{A}}$) decrease, as expected, owing to the plastic deformation of stress-concentrating asperities, as shown in Fig. 3. In Fig. 3, multiple points at each stress level correspond to the repeated profilometer scans and the mean values of the measurements are connected by the line. It should be noted that strain hardening of asperities is likely to have a moderating effect on this trend. Subjecting surfaces to multiple compression events brings about a marked decrease in surface roughness and amplitude, as shown in Fig. 4. Surfaces subjected to deformation at eight different interfacial configurations exhibited the lowest roughness and amplitude values of the samples tested. The effects of increased compression events are the result of new asperity–asperity couplings allowing

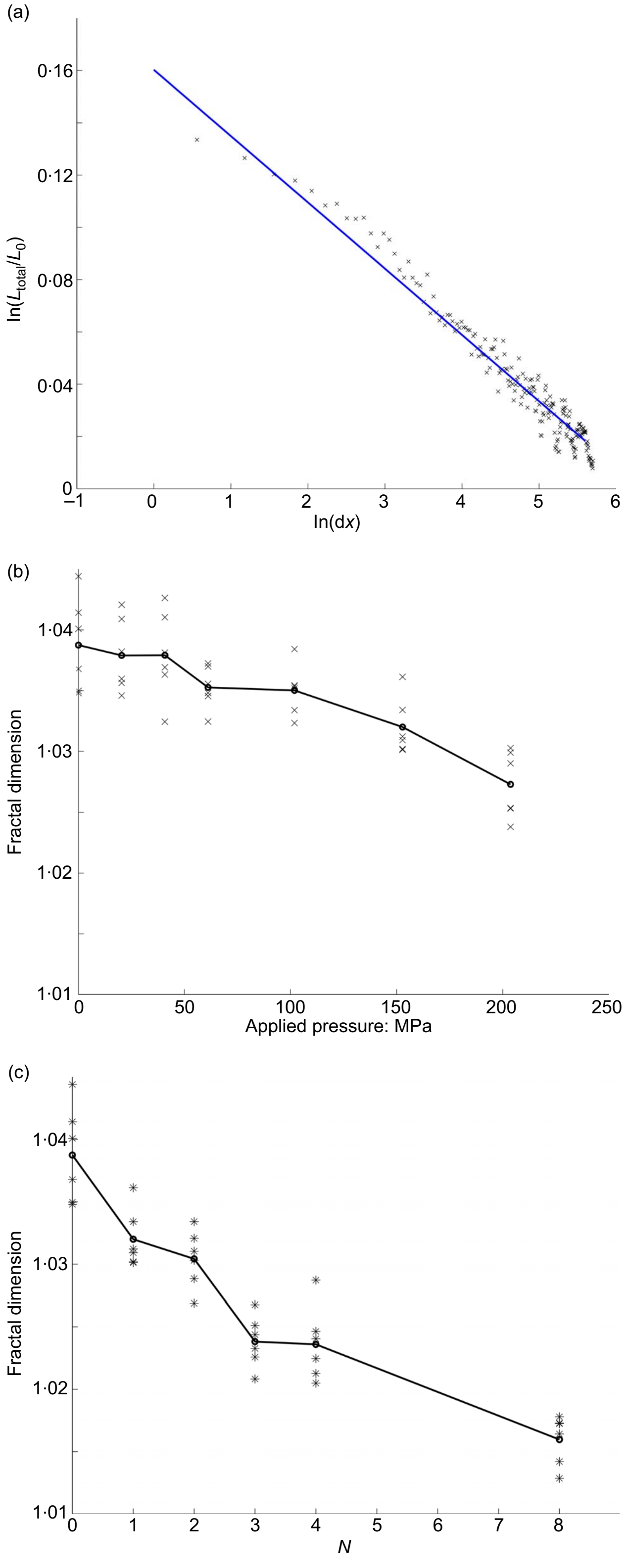

**Fig. 5.** (a) Measured profile length as a function of resolution used to determine fractal dimension. (b) Fractal dimension as a function of applied stress. (c) Fractal dimension as a function of number of loading events

the plastic deformation and breakage of a greater number of asperities.

The measurement of fractal dimension was carried out by analysis of the decrease in apparent profile length with coarsening measurement resolution following equation (3)

$$D = 1 - \frac{\Delta \log(L_{\text{total}}/L_0)}{\Delta \log(\mathrm{d}x)} \quad (3)$$

This is shown in Fig. 5(a). With increased stress or $N$, a decrease in fractal dimension occurs, as shown in Fig. 5(b) and 5(c). This is a consequence of decreasing roughness and profile flattening, consistent with other reported studies (Sun & Xu, 2005). Numerical values for fractal dimensions are lower than those reported elsewhere, as no scaling was applied to measured surface height values, which were significantly lower than the scan length ($L_0$ = 2000 μm).

The development of fractal dimension with surface deformation is illustrated in Fig. 6, where $\ln(D - 1)$ is plotted against the logarithm of normalised roughness $\ln(R_A/R_0)$ where $R_0$ = 1 μm. The greater linearity of the $\ln(D - 1)/\ln(R_A/R_0)$ distribution for samples subjected to multiple compression is an indication of deformation distributed across multiple scales. Despite an overall decreasing trend in both $D$ and $R_A$ in samples compressed in single events (Figs 3(a) and 5(b)), the lower $\ln(D - 1)/\ln(R_A/R_0)$ correlation found from individual surface scans seen in Fig. 6(a) is indicative of surface structure evolution occurring primarily through the deformation of higher level asperities with the retention of fine-scale roughness and consequently weak correlation of fractal dimension. Samples subjected to increasing load at a single compression event do not exhibit a significant change in static frictional coefficient at an interface with a rigid flat, as shown in Fig. 7. In contrast, samples compressed in multiple events showed a noticeable increase in adhesion to the quartz substrates.

### *Simulated fractal surfaces*

In Amonton–Coulomb friction, resistance to shear is independent of contact area. In the current model, friction is evaluated for contact of simulated fractal surfaces with a rigid flat plane from principles of atomic friction/adhesion, for which true contact area is a significant parameter. The fractal dimension of surfaces plays a role in determining the distribution of contact pressure across the surface structure and thus affects $\mu_S$ in adhesion-dominated interactions. In samples exhibiting higher fractal dimensions, contact stiffness is lower, resulting in greater asperity compliance as shown in Fig. 8(a). Furthermore, in samples exhibiting higher fractal dimensions, the number of contacting asperities $N_A$ and the total mean true contact area is higher for a given normal apparent contact stress, consequently $\sum_{i=1}^{i=N_A} \tau_0 A_i$ increases, where $A_i$ is the true contact area of individual asperities. For this reason, overall static friction is higher at higher initial $D$ values. From Fig. 8(b) it can be seen that samples of greater fractal dimensions exhibit higher static friction values, with the significance of this trend expectedly increasing for higher $\tau_0$ and lower $\alpha$ normalised values (lower $\alpha'/\tau_0'$ in the figure). $D$ values here refer to the fractal dimension of the initial unloaded simulated surface. With compression, the fractal dimension of simulated surfaces decreases, similar to the behaviour of experimental specimens, as shown in Fig. 9. Here, evolving $D$ was measured for a one-dimensional cross-section of the surface and deformation was applied to achieve a reduction of 20% in profile amplitude.

## DISCUSSION

The results show that the number of contact events plays an important role in the development of fractal surface structures and resultant frictional interactions. Roughened surfaces subjected to repeated loading were the only samples tested that exhibited a significant change in static friction. Surface analysis suggests that this may be the result of asperity deformation occurring across multiple scales of surface features, as evident from a greater corollary decrease in fractal dimension, resulting in an overall increased true contact area. It should be noted, however, that a similar overall trend of both decreasing fractal dimension and decreasing roughness was observed with both loading regimes.

Flattening tests were carried out on metallic specimens, which exhibit fundamentally different behaviour than brittle geomaterials, with the predominant mechanism of asperity flattening in the present work assumed to be plastic deformation (although asperity breakage cannot be excluded), which is appropriate for clay minerals. Nonetheless, the observations reported here imply that for geomaterials, particularly at high temperatures and stresses, the frequency of reorganisation events, which bring about new surface contact profiles, may govern the development of intergranular frictional forces. The fractal dimension of contacting surfaces plays a key role in governing the extent of frictional phenomena at the molecular level. In geomaterials, the time-dependent change of contact profiles and surface fractality due to phenomena of abrasion, erosion, fracture and plastic deformation may govern the evolution of intergranular friction and the development of force networks. The results illustrate the importance of evaluating surface structures and their development in order to appropriately predict the evolution of intergranular frictional strength.

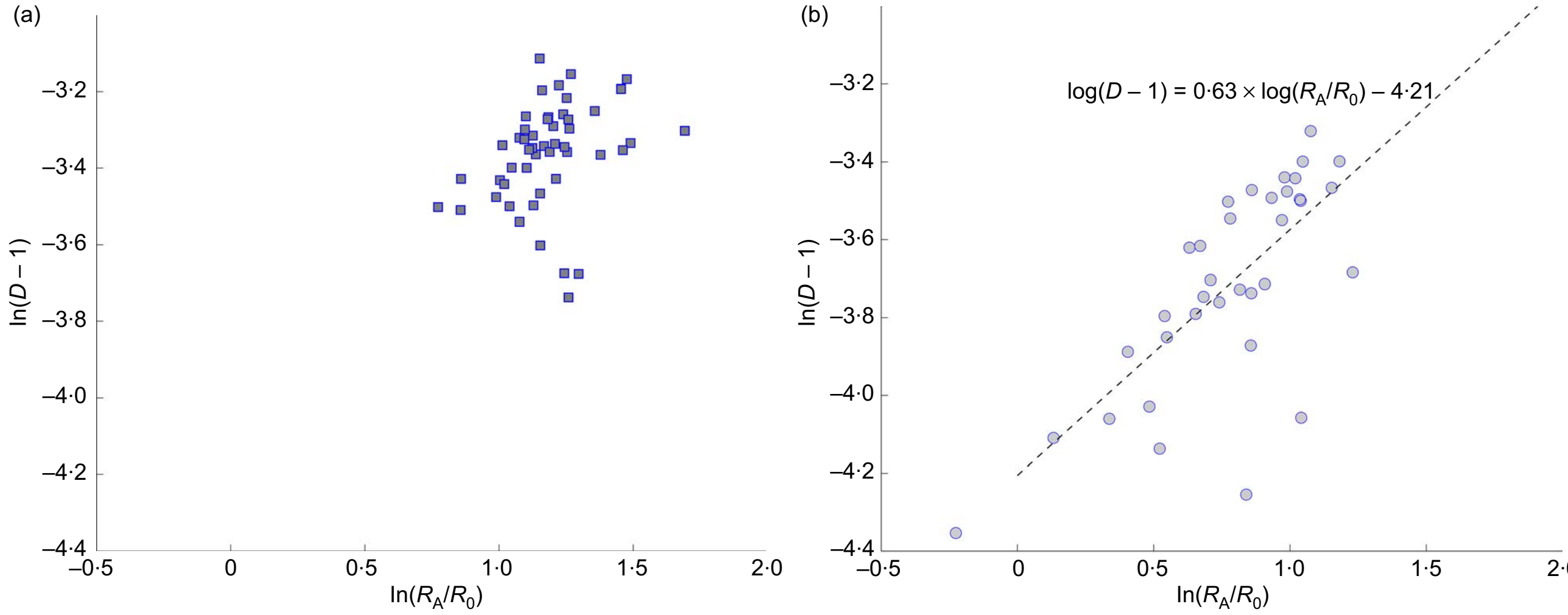

**Fig. 6.** Evolution of fractal dimension with changing roughness for (a) varied compressive load and (b) number of loading events. Points correspond to individual surface scans

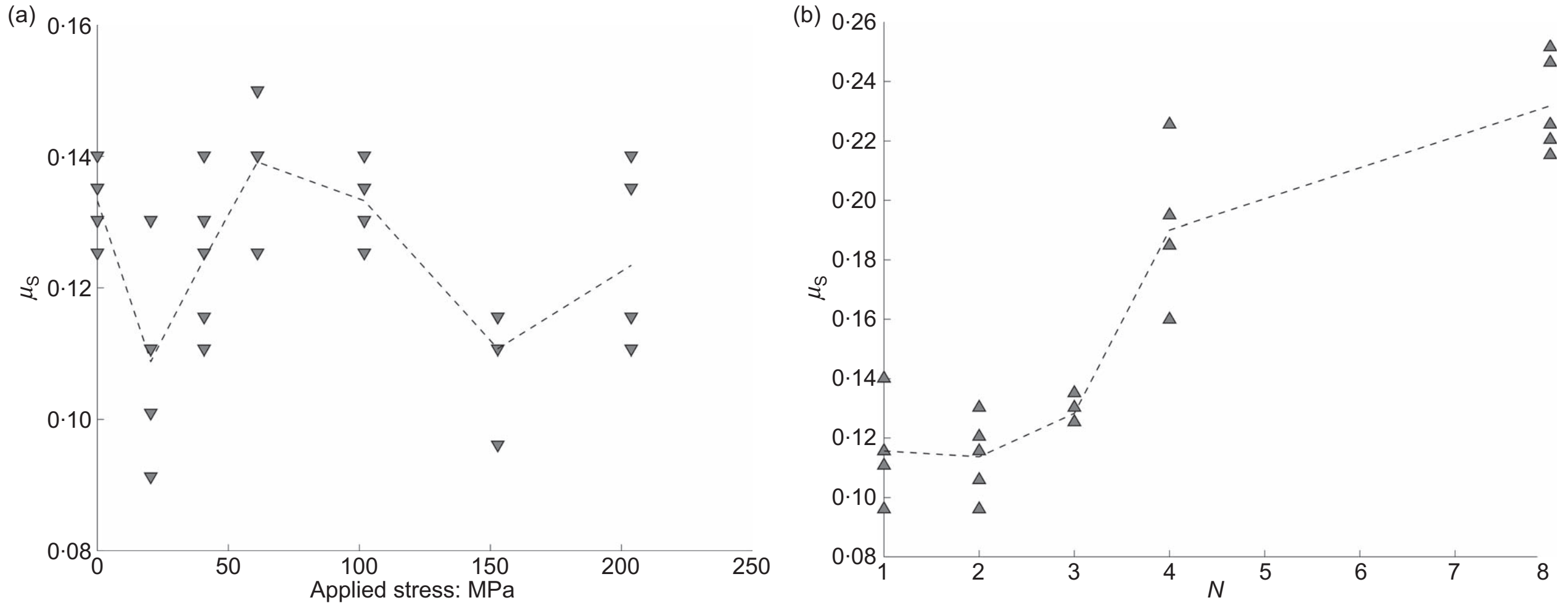

**Fig. 7.** Static friction at an interface with a quartz substrate for samples (a) compressed at varied loads and (b) varied number of loading events

From surface simulations it is evident that the significance of surface fractal dimension increases for adhesion-dominated frictional interactions. It is further evident that contact stiffness decreases with higher fractality in surface structures, with a larger displacement relative to the highest asperity peak occurring for a given applied stress. Static friction coefficients were found to be higher for simulated surfaces of higher fractal dimensions, in agreement with other reported studies (Bhushan *et al.*, 1995; Sun & Xu, 2005). While this observation is seemingly contradictory to the experimental results from the present work, it should be noted that the fractal dimensions of simulated surfaces were reported before surface deformation, while the fractal dimensions of experimental samples in the present work were assessed subsequent to compressive flattening of asperities. The development of fractal dimension of simulated surfaces with compression, shown in Fig. 9, reveals a similar proportional decrease to that observed in experimental samples, accompanying a decrease in profile amplitude of 20% (a similar amplitude reduction to that observed in samples). This illustrates the appropriateness of Weierstrass–Mandelbrot functions for the description of real surfaces.

Material properties and the relative dominance of asperity interactions, atomic friction coefficient and adhesion need to be considered in conjunction with the mode of surface structure deformation in order to determine the nature of the evolution of frictional forces at granular interfaces. Future work will investigate interactions of pairs of simulated rough fractal surfaces combining the contributions of physical asperity interlocking and molecular-level friction with an adhesive component. This will be supported by further experimental work investigating different modes of surface structure development and fractal evolution.

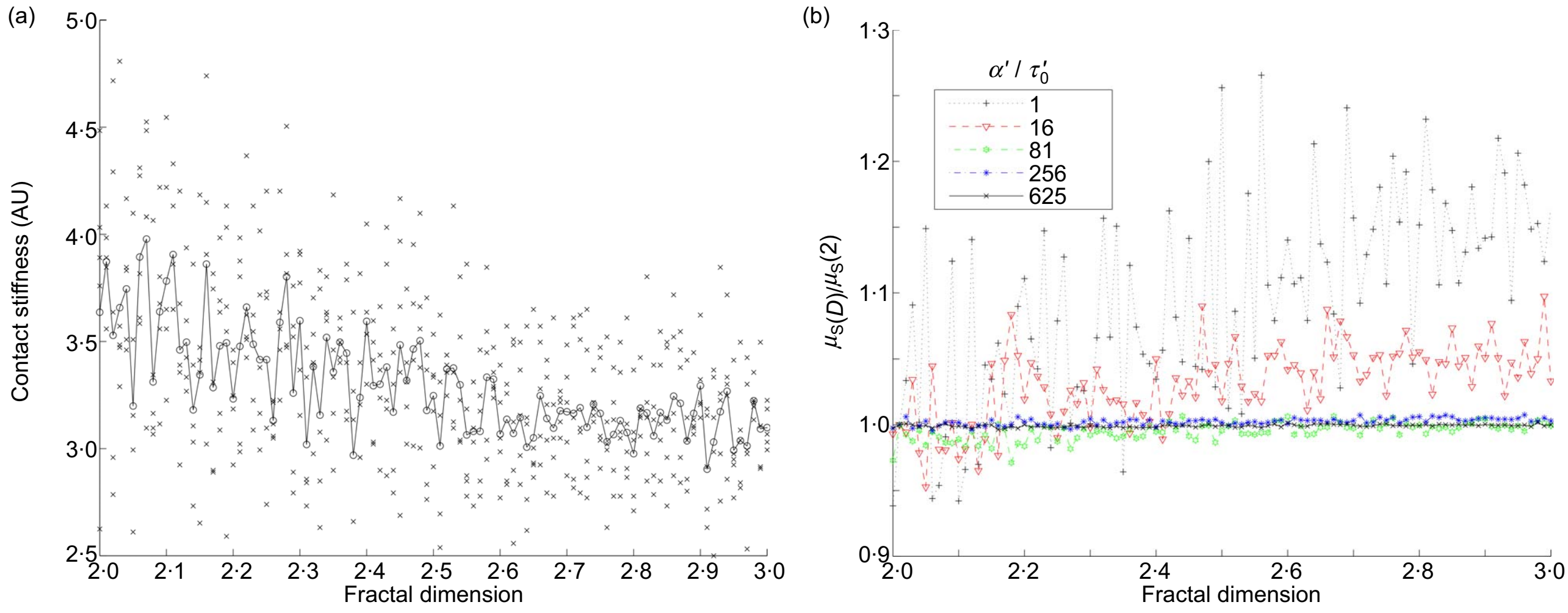

**Fig. 8.** Under constant simulated loading conditions, (a) contact stiffness for surfaces of different fractal dimensions and (b) normalised $\mu_S$ (relative to $\mu_S$ at $D = 2$) for simulated surfaces of different fractal dimensions with varied non-dimensionalised relative $\alpha/\tau_0$ values, applied to represent varied significance of adhesion/load-dependent component

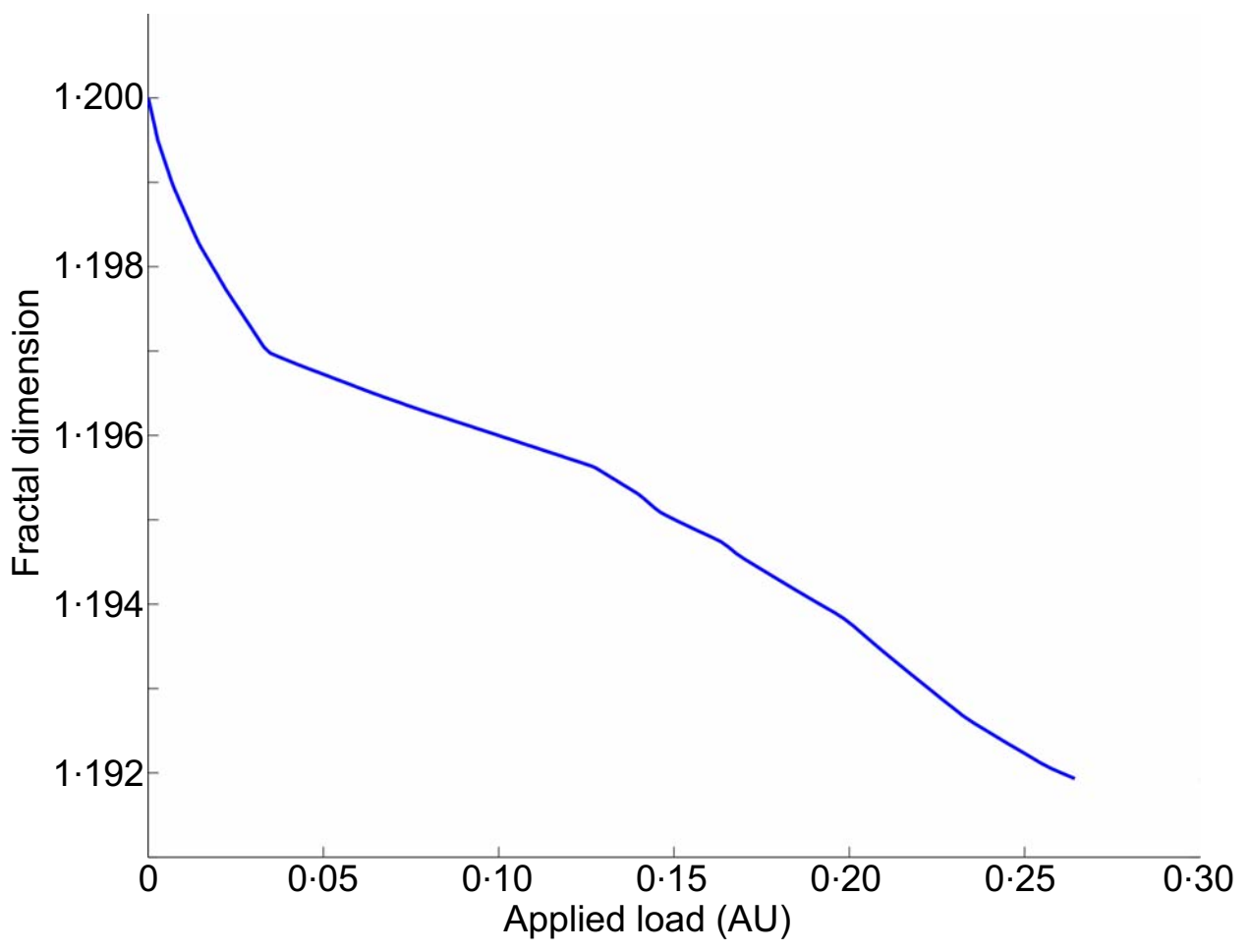

**Fig. 9.** Decrease in fractal dimension of a simulated surface with increasing compression against a rigid flat counter surface

## Acknowledgement

Financial support for this research from the Australian Research Council through grant no. DP120104926 is gratefully appreciated.

## REFERENCES

Berry, M. & Lewis, Z. (1980). On the Weierstrass–Mandelbrot fractal function. *Proc. Royal Soc. London A Math. Phys. Sci.* **370**, No. 1743, 459–484.

Bhushan, B., Israelachvili, J. N. & Landman, U. (1995). Nanotribology: friction, wear and lubrication at the atomic scale. *Nature* **374**, No. 6523, 607–616.

Boitnott, G., Biegel, R., Scholz, C., Yoshioka, N. & Wang, W. (1992). Micromechanics of rock friction 2: Quantitative modeling of initial friction with contact theory. *J. Geophys. Res.* **97**, No. B6, 8965–8978.

Bowden, F. P. & Tabor, D. (2001). *The friction and lubrication of solids*. New York: Oxford University Press.

Buzio, R., Boragno, C., Biscarini, F., De Mongeot, F. B. & Valbusa, U. (2003). The contact mechanics of fractal surfaces. *Nature Mater.* **2**, No. 4, 233–237.

Cavarretta, I., Coop, M. & O'Sullivan, C. (2010). The influence of particle characteristics on the behaviour of coarse grained soils. *Géotechnique* **60**, No. 6, 413–423.

Deng, Z., Smolyanitsky, A., Li, Q., Feng, X. Q. & Cannara, R. J. (2012). Adhesion-dependent negative friction coefficient on chemically modified graphite at the nanoscale. *Nature Mater.* **11**, No. 12, 1032–1037.

Goedecke, A., Jackson, R. & Mock, R. (2013). A fractal expansion of a three dimensional elastic–plastic multi-scale rough surface contact model. *Tribology Int.* **59**, 230–239.

Greenwood, J. & Williamson, J. (1966). Contact of nominally flat surfaces. *Proc. Royal Soc. London A Math. Phys. Sci.* **295**, No. 1442, 300–319.

Harrison, J., White, C., Colton, R. & Brenner, D. (1992). Molecular-dynamics simulations of atomic-scale friction of diamond surfaces. *Phys. Rev. B* **46**, No. 15, 9700–9708.

Hasegawa, M., Liu, J., Okuda, K. & Nunobiki, M. (1996). Calculation of the fractal dimensions of machined surface profiles. *Wear* **192**, No. 1, 40–45.

Li, Q. & Kim, K. S. (2008). Micromechanics of friction: effects of nanometre-scale roughness. *Proc. Royal Soc. London A Math. Phys. Sci.* **464**, No. 2093, 1319–1343.

Majumdar, A. & Tien, C. (1990). Fractal characterization and simulation of rough surfaces. *Wear* **136**, No. 2, 313–327.

Patra, S., Ali, S. & Sahoo, P. (2008). Elastic-plastic adhesive contact of rough surfaces with asymmetric distribution of asperity heights. *Wear* **265**, No. 3–4, 554–559.

Persson, B. N. J. (2006). Contact mechanics for randomly rough surfaces. *Surf. Sci. Rep.* **61**, No. 4, 201–227.

Persson, B. (2007). Relation between interfacial separation and load: a general theory of contact mechanics. *Phys. Rev. Lett.* **99**, No. 12, 1–4.

Sadrekarimi, A. & Olson, S. (2011). Critical state friction angle of sands. *Géotechnique* **61**, No. 9, 771–783.

Sammis, C. G. & Biegel, R. L. (1989). Fractals, fault-gouge, and friction. *Pure Appl. Geophys.* **131**, No. 1, 255–271.

Sun, D. & Xu, Y. (2005). Correlation of surface fractal dimension with frictional angle at critical state of sands. *Géotechnique* **55**, No. 9, 691–696.

Yan, W. & Komvopoulos, K. (1998). Contact analysis of elastic-plastic fractal surfaces. *J. Appl. Phys.* **84**, No. 7, 3617–3624.